\documentclass[aps,prb,amsmath,amssymb,superscriptaddress,twocolumn,10pt]{revtex4-2}
\usepackage{braket}
\usepackage{graphicx}
\usepackage{hyperref}
\usepackage{bbm,color,ulem}
\DeclareMathAlphabet{\mathpzc}{OT1}{pzc}{m}{it} 
\begin{document}
\title{Crosstalk- and charge-noise-induced multiqubit decoherence in exchange-coupled quantum dot spin qubit arrays}
\author{Robert E.~Throckmorton}
\author{S.~\surname{Das Sarma}}
\affiliation{Condensed Matter Theory Center and Joint Quantum Institute, Department of Physics, University of Maryland, College Park, Maryland 20742-4111 USA}
\date{\today}
\begin{abstract}
We determine the interqubit crosstalk- and charge-noise-induced decoherence time $T_2^\ast$ for a system of $L$ exchange-coupled electronic spin qubits in arrays of size $L=3$--$14$ for a number of different multiqubit geometries by directly calculating the return probability.  We compare the behavior of the return probability to other quantities, namely, the average spin, the Hamming distance, and the entanglement entropy.  In all cases, we use a starting state with alternating spins, $\ket{\Psi_0}=\ket{\downarrow\uparrow\downarrow\cdots}$.  We show that a power law behavior, $T_2^\ast\propto L^{-\gamma}$, is a good fit to the results for the chain and ring geometries as a function of the number of qubits, and provide numerical results for the exponent $\gamma$.  We find that $T_2^\ast$ depends crucially on the multiqubit geometry of the system.  We also calculate the expectation value of one of the spins, the Hamming distance, and the entanglement entropy and show that they are good proxies for the return probability for measuring $T_2^\ast$.  A key finding is that $T_2^\ast$ decreases with increasing $L$.  We also demonstrate that these results may be understood in terms of perturbation theory and its breakdown.
\end{abstract}
\maketitle

\section{Introduction}
In order to develop a working quantum computer, one must be able to combine many qubits and achieve sufficiently high fidelity in both one- and two-qubit operations, with $99.9\%$ being the minimum required fidelity to employ surface code techniques for error correction.  In particular, Si spin qubits offer the greatest hope for scalability due to their compatibility with the existing Si electronics technology and their small size compared to other qubits, but currently have lower fidelities than competing qubit platforms such as ion traps and superconducting junctions.  Experiments have considered single-  \cite{s41928-019-0234-1,acs.nanolett.0c02397,acs.nanolett.0c04771} and two-qubit gates \cite{nnano.2014.216,nature15263,s41586-019-1197-0.2019,science.aao5965,nature25766,PhysRevX.9.021011,arxiv.2111.11937,s41586-021-04182-y,s41586-021-04273-w} in both single- and multiqubit systems, achieving single-qubit gate fidelities as high as $99.9\%$, and three experiments in particular \cite{arxiv.2111.11937,s41586-021-04182-y,s41586-021-04273-w} have demonstrated two-qubit gate fidelities near or, in some cases, even exceeding $99.9\%$.  We thus see that experiments on semiconductor-based spin qubits are very close to achieving the $99.9\%$ fidelity in both single- {\it and} two-qubit gates that is required in order to implement error correction techniques.

One major source of errors in gate operations is decoherence, which results in loss of information in the qubit system over time.  This decoherence is characterized by the decoherence time $T_2^\ast$, which we will define here as the time scale over which the return probability $P_R(t)$, or the probability that the system will be measured to be in its initial state $\ket{\Psi_0}$ at time $t$,
\begin{equation}
    P_R(t)=\left |\braket{\Psi_0|\Psi(t)}\right |^2, \label{Eq:ReturnP}
\end{equation}
decays to its long-time steady-state value.  Experiments that measure dephasing times, or the time scale over which the expectation value of a spin decays to its steady-state value, both with and without the use of sequences such as Hahn echo and Carr-Purcell-Meiboom-Gill (CPMG), have been undertaken, finding dephasing times as high as $0.12$ ms without these special sequences and as high as $28$ ms with these sequences \cite{nature15263,PhysRevLett.108.086802,nnano.2014.216,PhysRevLett.110.146804,s41565-017-0014-x,nnano.2016.170}.  While the dephasing time is not the same as the decoherence time that we will consider here, these two times are correlated with one another due to being controlled by the same factors, so that any improvement in decoherence time will translate into an improvement in dephasing time as well.

There are two effects that contribute to this decoherence time.  One is noise in the couplings between qubits, which comes from voltage fluctuations in the gates defining the quantum dots and from charged defects in the semiconductor itself.  This will, in fact, be the main source of noise-induced decoherence in Si because it may be isotopically purified to remove ${}^{29}$Si, the lone stable magnetic isotope of Si.  This almost completely eliminates Overhauser noise, which would manifest as an effective fluctuating magnetic field applied to the electronic spins, adding another source of noise to the system.  We ignore Overhauser noise in the current work since it is unimportant for the experimental Si spin qubits.  Another decohering mechanism is the interactions between qubits, which, while necessary to perform two-qubit gates, also introduce crosstalk, which is detrimental to single-qubit gates and to the ability of the individual qubits to retain their states in the absence of gate operations.  We expect these interaction effects to depend on the exact arrangement of the qubits, due simply to the different connectivities for different multiqubit geometries.  Higher connectivity of a qubit to other qubits helps with performing two-qubit gates between two arbitrary qubits, but it can also exacerbate crosstalk due to direct interaction of one qubit with a larger number of other qubits.

It is thus critical to study the effects of noise, the number of qubits, and the arrangement of the qubits on the decoherence time, as they will inform the design of a multiqubit quantum computer circuit.  Theoretical studies of decoherence in semiconductor spin qubit systems so far have focused on smaller systems, up to four spins \cite{PhysRevB.95.085405,PhysRevB.103.205402,s41534-022-00523-5}.  While experiments on semiconductor-based electron spin qubits so far have only small numbers of qubits (at most four), it is important to study larger systems theoretically because experiments will eventually need to scale up to larger systems as well.  We thus extend these analyses to larger systems, and to other multiqubit geometries that are possible with such larger system sizes.

We study a system of $L$ spin qubits with nearest-neighbor Heisenberg exchange couplings in a number of different multiqubit geometries, including chains, rings, and a number of two-dimensional arrays.  We include quasistatic charge noise in the exchange couplings, modeled as a Gaussian distribution with mean $J_0$ and standard deviation $\sigma_J$.  Charge noise fluctuations observed in experiments have frequencies on the order of megahertz (i.e., timescales on the order of $\mu$s), while typical gate operations in semiconductor-based spin qubit systems have durations on the order of nanoseconds, so we expect quasistatic charge noise to approximate the effects of noise in real experimental systems very well.

We begin by numerically calculating the return probability as a function of time using an alternating arrangement of spins, $\ket{\Psi_0}=\ket{\downarrow\uparrow\downarrow\cdots}$, as the initial state.  We then extract the decoherence time $T_2^\ast$ from this return probability by finding the best empirical fit to an envelope of the form,
\begin{equation}
    E(t)=P_\infty\pm(1-P_\infty)e^{-(t/T_2^\ast)^\alpha}. \label{Eq:EnvFittingForm}
\end{equation}
We provide plots of our results for the chain and ring geometries as a function of $L$, as well as a table of all of our results, for four values of $\sigma_J$.  We find that our results for the chain and ring geometries fit power laws, i.e., $T_2^\ast\propto L^{-\gamma}$, very well, and we report our results for the exponent $\gamma$ for several values of $\sigma_J$.  We find that, for the ring geometry, $\gamma$ tends to decrease with increasing $\sigma_J$, while, for the chain geometry, $\gamma$ instead trends upward.  We also compare different multiqubit geometries for a given number of qubits, and find that $T_2^\ast$ can vary greatly for fixed $L$ depending on the geometry, sometimes even by an order of magnitude.  Overall, we see that $T_2^\ast$ decreases with increasing $L$, regardless of the specific geometry considered.  We also provide the values of $\alpha$ obtained from this fit; we find that, for small system sizes (at most four spins), $\alpha=2$, and thus the decay has a Gaussian profile, as found in the previous work.  However, for larger system sizes, smaller values of $\alpha$ are found, thus indicating that the decay of the system is no longer Gaussian for such large systems, and may be approaching an exponential decay.  We note that a similar work considering just two qubits \cite{s41534-022-00523-5} found similar results, with the only difference between the Gaussian and exponential regimes being the strength of the noise; there was no fundamental difference in its nature.

We show that these results may be understood within perturbation theory.  We derive an expression for the noise-averaged return probability at first order in the noise-induced deviations of the exchange couplings from their intended value, showing that we obtain a sum of sinusoidal oscillations with Gaussian damping.  We point out two ways, using this expression, that one can find deviations from Gaussian damping.  The first is, assuming that perturbation theory holds true, that one may see a large number of these terms, all with different damping rates; such a large number of terms will yield an overall non-Gaussian decay profile.  The other is that perturbation theory simply fails, and thus we cannot claim even approximately Gaussian decay.  Both of these tend to occur for large system sizes and/or large $\sigma_J$, consistent with our results.  We find, however, that the first of these, the proliferation of terms with increasing system size, is by far the dominant factor; we see that perturbation theory works surprisingly well, even for larger systems, for which we would expect it to break down even for the smallest values of $\sigma_J$ that we consider.  Ultimately, this proliferation of terms is simply due to the fact that, as the system size grows, the number of energy eigenstates within the $S_z=0$ (even $L$) or $S_z=-\tfrac{1}{2}$ (odd $L$) subspace grows exponentially.  This results in a rapidly increasing number of possible level transitions, which in turn gives a larger number of terms in the perturbative expansion, even at first order.

In addition to calculating the return probability, it is helpful to illustrate the effects of decoherence in other quantities as well.  We thus calculate the average of one of the spins in the system, the Hamming distance, and the entanglement entropy.  The first two in particular are measurable in experiments, making their calculation an especially valuable guide for future experiments, and we show that they can serve as suitable proxies for the return probability for determining $T_2^\ast$.

We should emphasize here that our results do not describe experiments that have already been performed; rather, our reason for discussing the variation of $T_2^\ast$ with system size and geometry is to inform future experiments, even in experiments in which gates are performed.  There is a correlation between $T_2^\ast$, which describes the behavior of the system if it is left to evolve on its own, and how long a quantum computation can take (and thus the number of gates that can be performed) before there is an unacceptable loss of information due to the fact that the same factors affect both.  Therefore, it is critical that we study $T_2^\ast$, as it will inform the design of future experiments, regardless of whether or not gates are performed.

The rest of the paper is organized as follows.  We introduce our model in Sec.~\ref{sec:Model}.  We numerically calculate exactly the return probability as a function of time for the ring and chain geometries, as well as for a number of two-dimensional arrays, and extract the decoherence time $T_2^\ast$ from it in Sec.~\ref{sec:RTandT2S}.  We explain how the change from Gaussian decay to non-Gaussian decay that we find can be understood through perturbation theory and its breakdown in Sec.~\ref{sec:Analysis}.  We determine other quantities---average spin, Hamming distance, and entanglement entropy---in Sec.~\ref{sec:OtherQs}.  Finally, we give our conclusions in Sec.~\ref{sec:Conclusion}.

\section{Model} \label{sec:Model}
We consider here a system of $L$ spin qubits with nearest-neighbor Heisenberg exchange couplings:
\begin{equation}
H=\sum_{\braket{ij}}J_{ij}\vec{\sigma}_i\cdot\vec{\sigma}_j,
\end{equation}
where $\vec{\sigma}_i$ is the vector of Pauli matrices acting on spin $i$.  This Hamiltonian with suitable $J_{ij}$ values describes the appropriate spin qubit coupling in most Si quantum dot qubit platforms being used currently.  We consider several different geometries for the system.  For all system sizes, we consider both chain and ring geometries and additionally consider other, two-dimensional geometries for certain system sizes, as will be described later.  We include quasistatic noise in the exchange couplings $J_{ij}$, modeled as a Gaussian distribution with mean $J_0$ and standard deviation $\sigma_J$,
\begin{equation}
f_J(J)\propto e^{-(J-J_0)^2/2\sigma_J^2}.
\end{equation}
In all calculations, we use as our starting state an alternating arrangement of spins, $\ket{\Psi_0}=\ket{\downarrow\uparrow\downarrow\cdots}$, which resides in the $S_z=0$ subspace for even $L$ and in the $S_z=-\tfrac{1}{2}$ subspace for odd $L$.  We use $20000$ realizations of disorder, numerically exactly solving the Schr\"odinger equation for each realization, and average the results for all quantities that we calculate here.  We consider system sizes of $L=3$--$14$ spin qubits.

\section{Return probability and decoherence time} \label{sec:RTandT2S}
\begin{figure}
	\centering
		\includegraphics[width=\columnwidth]{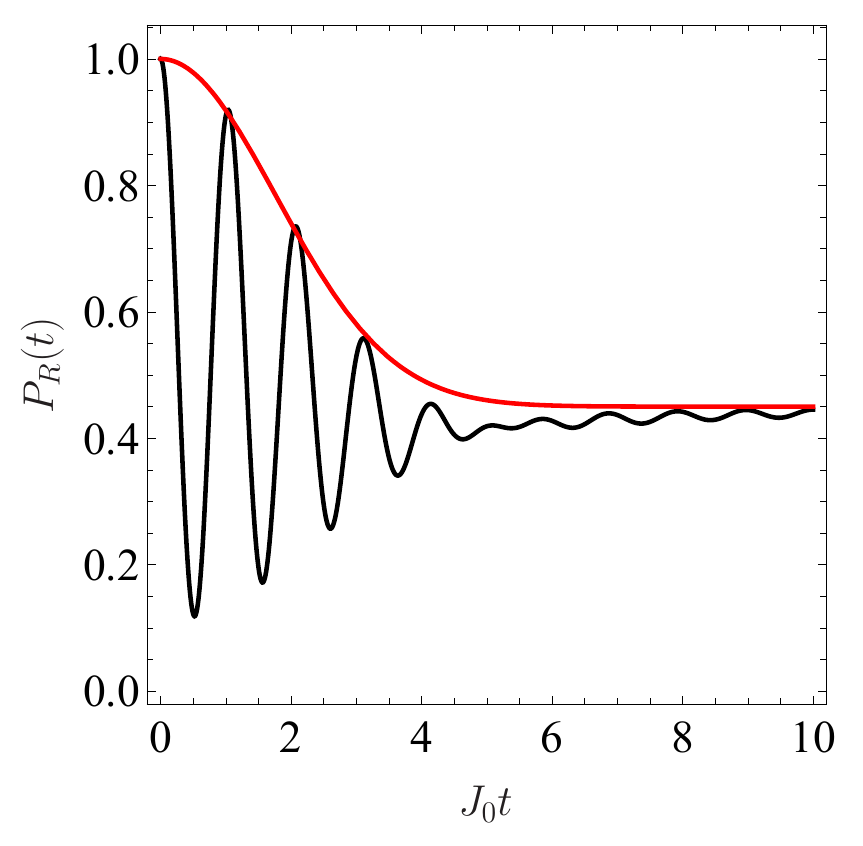}
	\caption{Example of fitting of an envelope of the form of Eq.~\eqref{Eq:EnvFittingForm}, for $L=3$ and the ring geometry.}
	\label{fig:FitExample}
\end{figure}

\begin{figure}
	\centering
		\includegraphics[width=0.49\columnwidth]{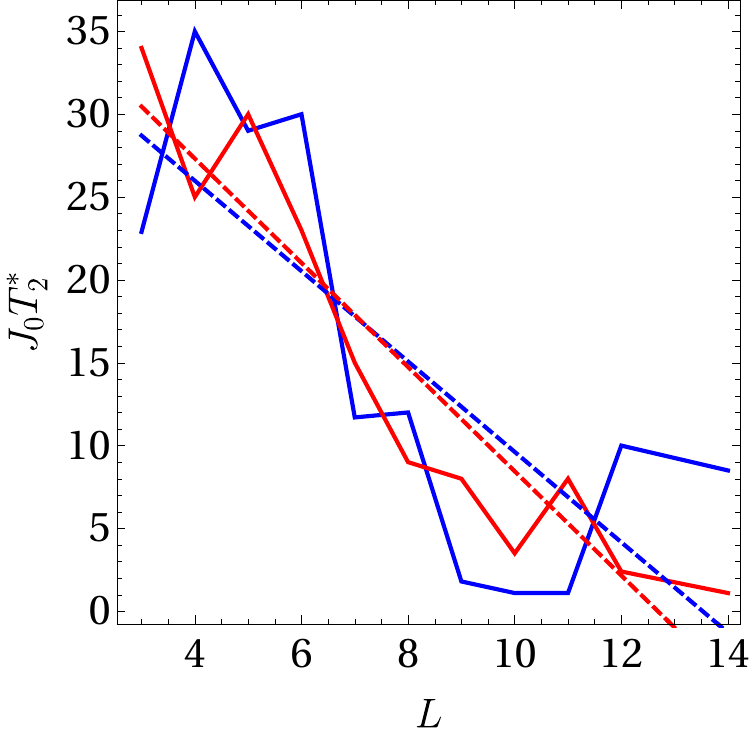}
		\includegraphics[width=0.49\columnwidth]{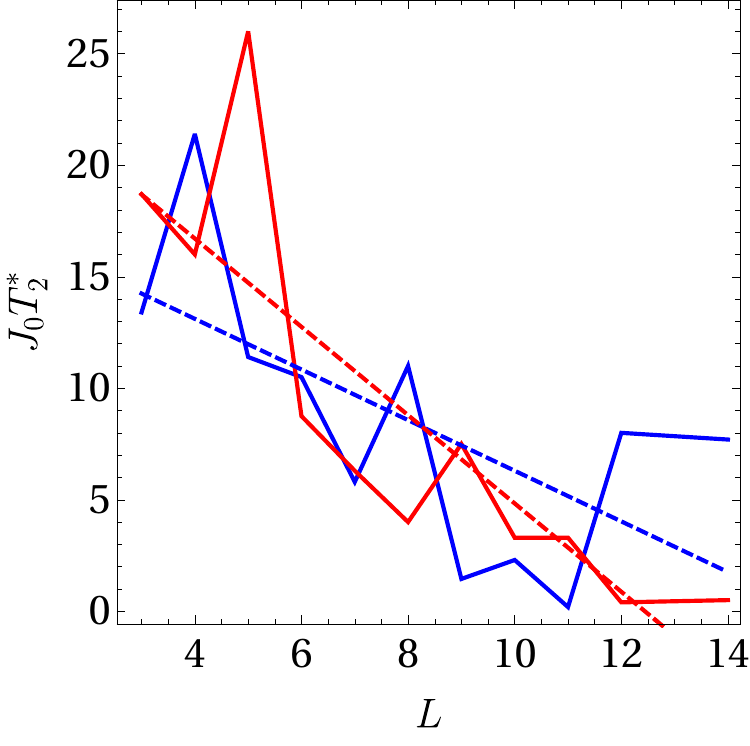}
		\includegraphics[width=0.49\columnwidth]{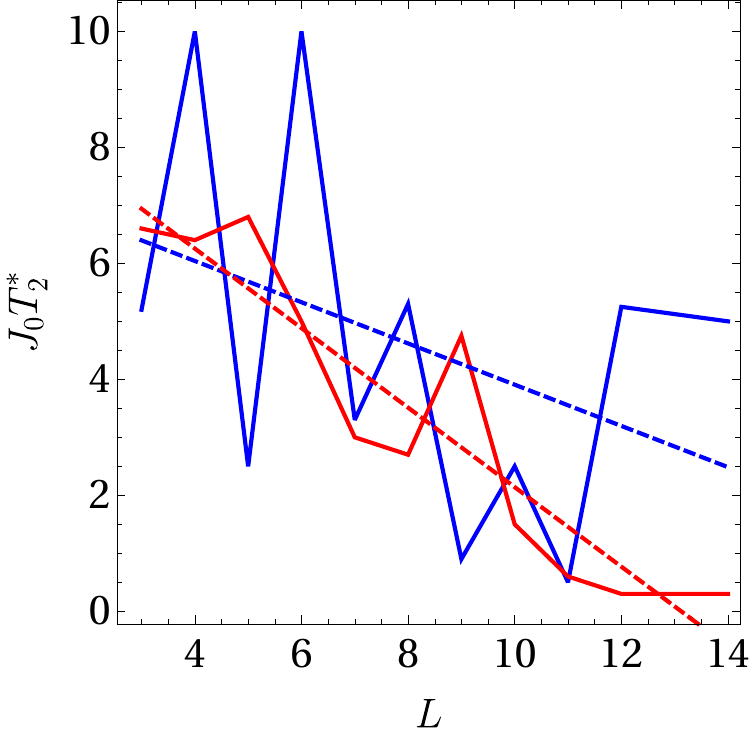}
		\includegraphics[width=0.49\columnwidth]{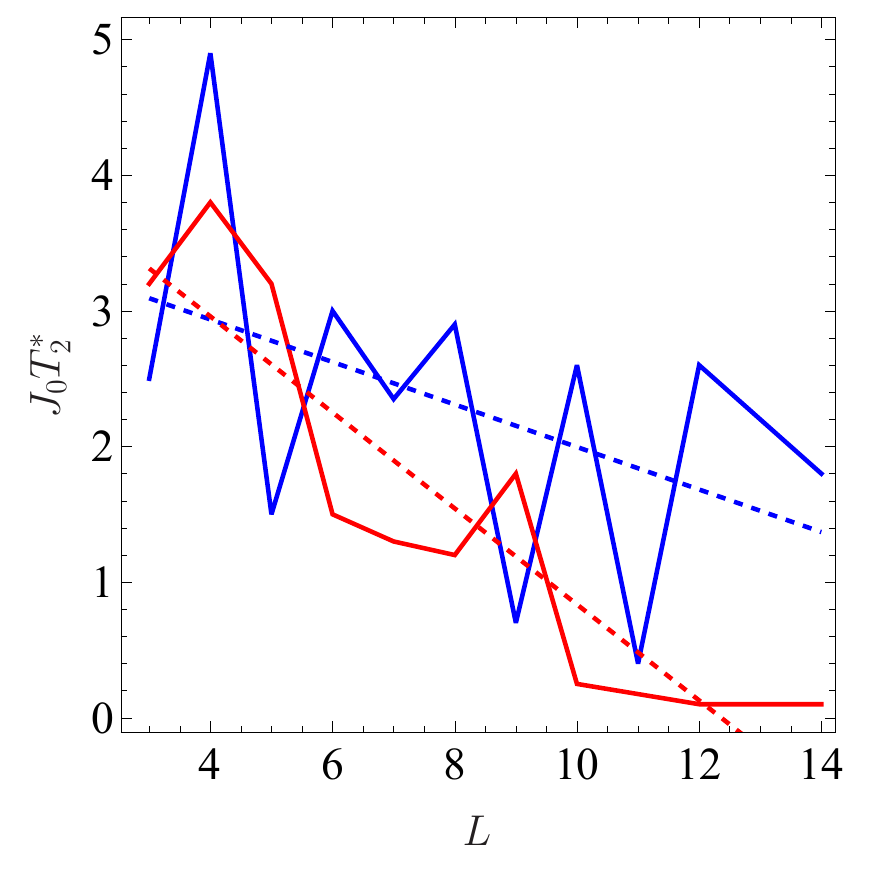}
	\caption{Plot of $T_2^\ast$ as a function of $L$ for $L=3$--$14$ and for $\sigma_J=0.01J_0$	(top left), $\sigma_J=0.02J_0$ (top right), $\sigma_J=0.05J_0$ (bottom left), and $\sigma_J=0.1J_0$ (bottom right) for ring (blue) and chain (red) geometries.  The dashed lines are linear fits, provided as a guide to the eye.}
	\label{fig:T2SPlots}
\end{figure}

\begin{figure}
	\centering
		\includegraphics[width=0.49\columnwidth]{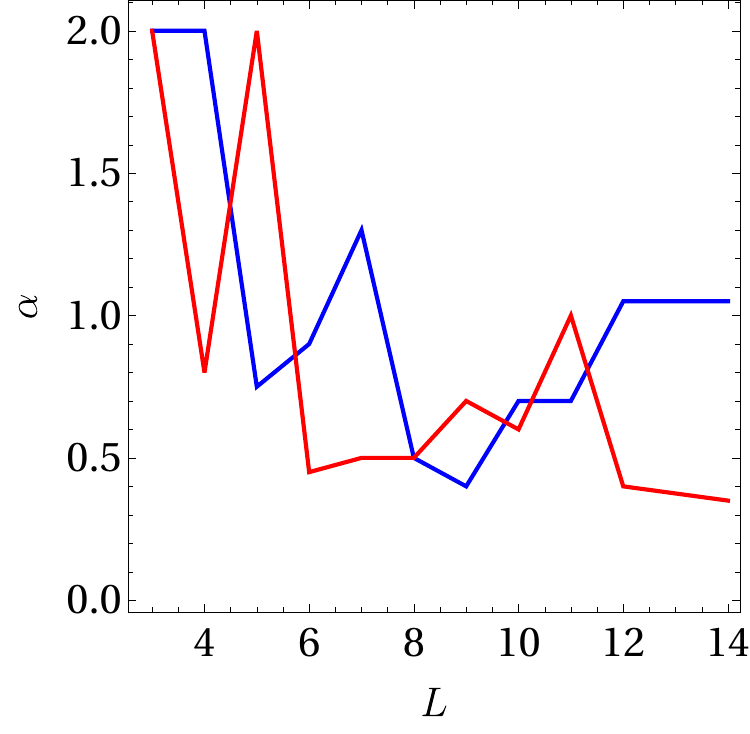}
		\includegraphics[width=0.49\columnwidth]{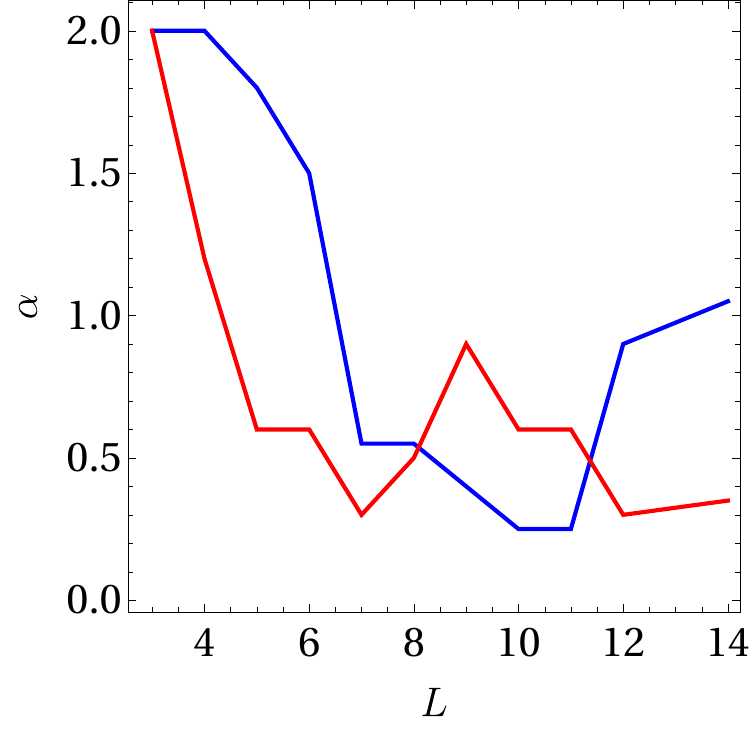}
		\includegraphics[width=0.49\columnwidth]{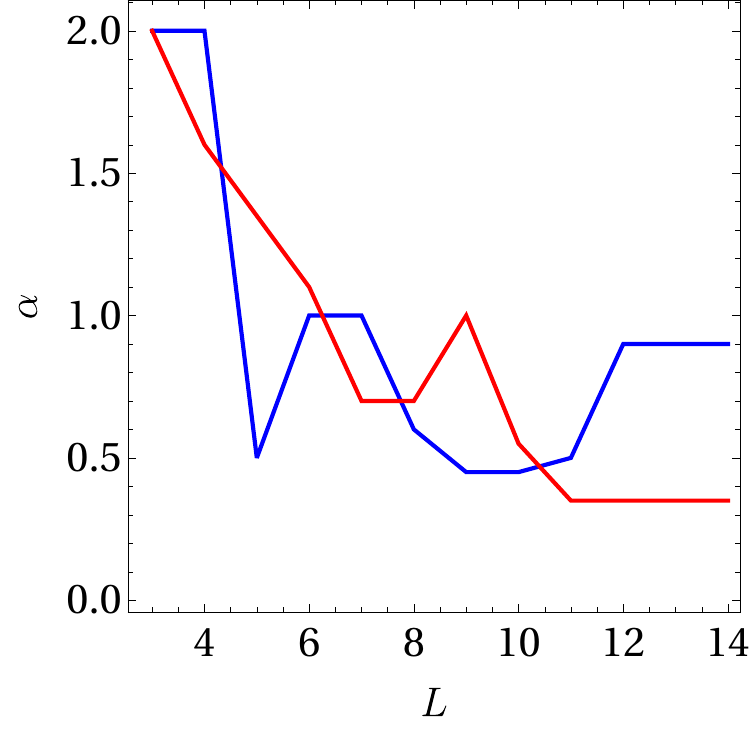}
		\includegraphics[width=0.49\columnwidth]{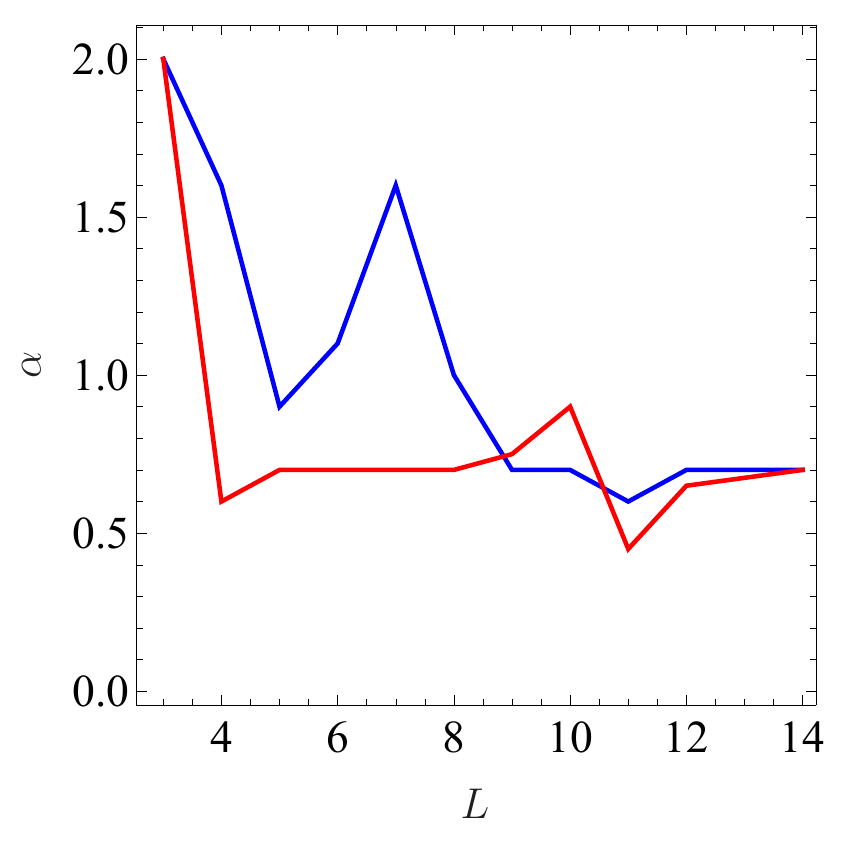}
	\caption{Plot of $\alpha$ as a function of $L$ for $L=3$--$14$ and for $\sigma_J=0.01J_0$ (top left), $\sigma_J=0.02J_0$ (top right), $\sigma_J=0.05J_0$ (bottom left), and $\sigma_J=0.1J_0$ (bottom right) for ring (blue) and chain (red) geometries.}
	\label{fig:AlphaPlots}
\end{figure}

\begin{figure}
	\centering
		\includegraphics[width=0.49\columnwidth]{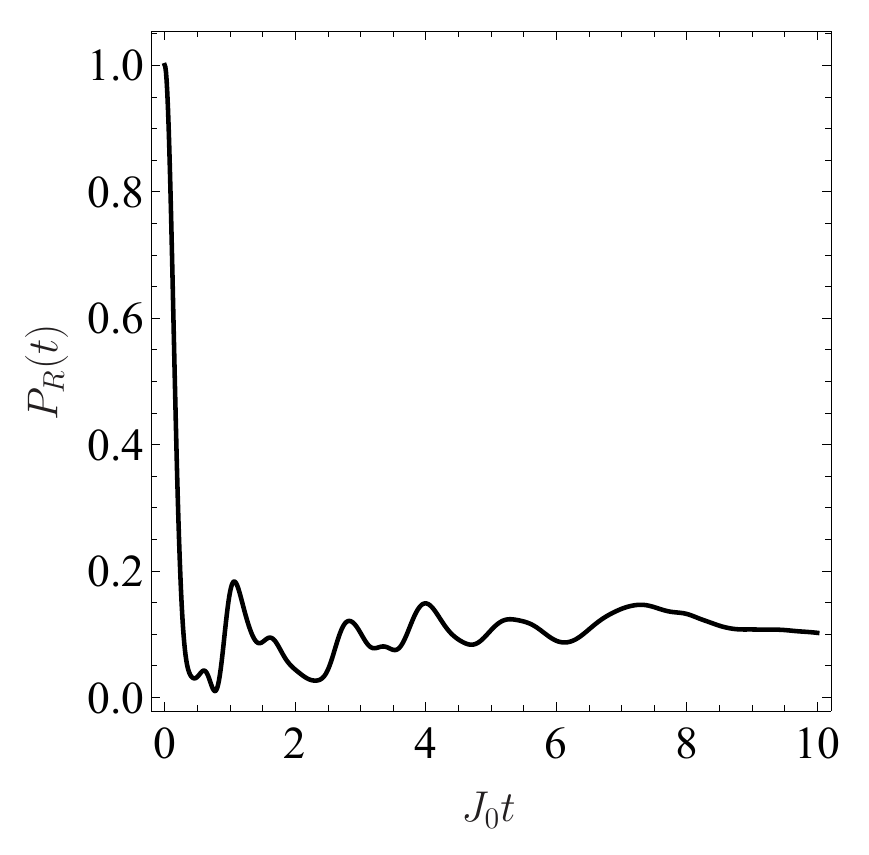}
		\includegraphics[width=0.49\columnwidth]{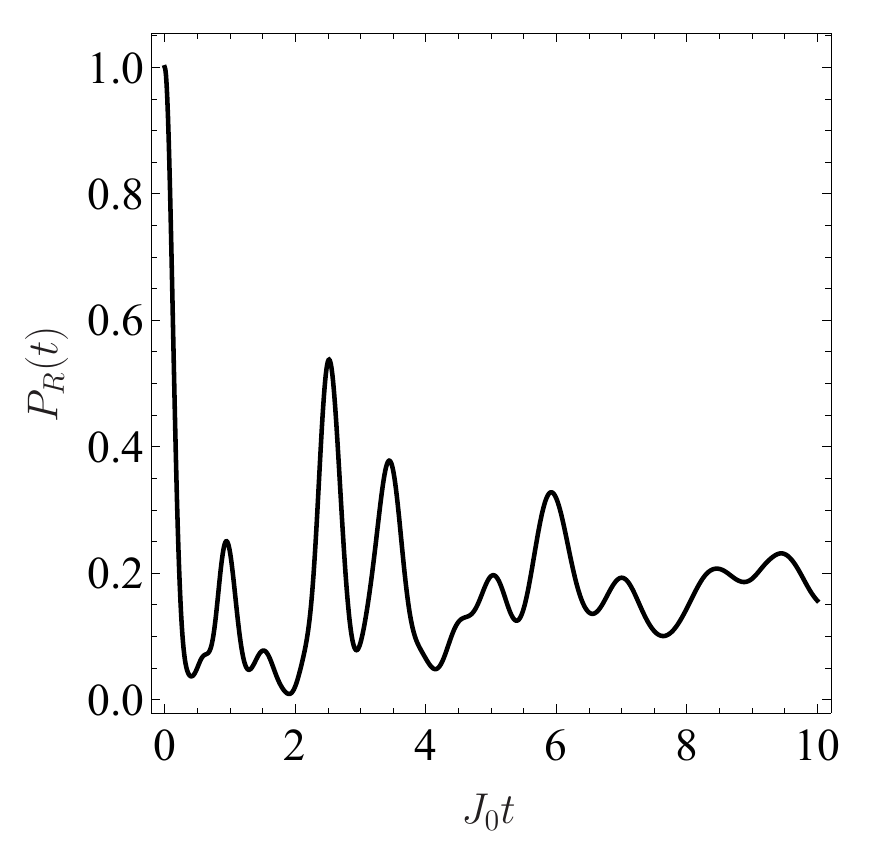}
	\caption{Plot of return probability $P_R(t)$ as a function of time for $L=8$ for chain (left) and ring (right) geometries.}
	\label{fig:RP_8Spins_Plot}
\end{figure}

The first quantity that we consider is the return probability, given by Eq.~\eqref{Eq:ReturnP}.  We calculate this probability as a function of time as described above.  We then determine the decoherence time $T_2^\ast$ as a function of $L$ for both the chain and ring geometries.  We determine this time by visually determining the curve of the form given in Eq.~\eqref{Eq:EnvFittingForm} that best approximates the envelope of $P_R(t)$.  We provide an illustration of this in Fig.~\ref{fig:FitExample} for a three-spin ring, and present the results for all $L=3$--$14$, for $\sigma_J=0.01J_0$, $0.02J_0$, $0.05J_0$, and $0.1J_0$, and for both the chain and ring geometries in Fig.~\ref{fig:T2SPlots}.  We note here that our results for the $L=3$ and $4$ chains and rings are consistent with the previous work.  Our numbers differ due to us absorbing a factor of $\tfrac{1}{4}$ into the exchange couplings that Ref.~\cite{PhysRevB.103.205402} did not; that is, the difference is simply due to a different definition of the exchange couplings $J_{ij}$.  We also present plots of the corresponding $\alpha$ values in Fig.~\ref{fig:AlphaPlots}.
We see that the behavior of $T_2^\ast$ is different between the two geometries.  To illustrate this visually, we show plots of $P_R(t)$ for $L=8$ in both the chain and ring geometries in Fig.~\ref{fig:RP_8Spins_Plot}.
We now fit these results to power laws, $J_0T_2^\ast(L)=\tau_0L^{-\gamma}$, via least-squares regression.  We do this for $\sigma_J=0.01J_0$, $0.02J_0$, $0.05J_0$, and $0.1J_0$.  We present the results for the exponent $\gamma$ in Table \ref{tab:GammaValues}.  From these values of $\gamma$, we find that the power law exponent shows a downward trend with increasing $\sigma_J$ for the ring geometry but shows an upward trend for the chain geometry.
\begin{table*}
\centering
		\begin{tabular}{|c|c|c|}
			\hline
			{$\sigma_J/J_0$} & {$\gamma$ (ring)} & {$\gamma$ (chain)} \\
			\hline
			\hline
            0.01 & $1.78529\pm 0.680195$ & $2.08751\pm 0.333944$ \\
			\hline
			0.02 & $1.49418\pm 0.787218$ & $2.40919\pm 0.483226$ \\
			\hline
			0.05 & $0.762535\pm 0.591355$ & $2.14082\pm 0.444261$ \\
			\hline
			0.1 & $0.607596\pm 0.449304$ & $2.58247\pm 0.464581$ \\
			\hline		
		\end{tabular}
	\caption{Exponents of power law fits to $T_2^\ast$ as a function of $L$ for different values of $\sigma_J$.  The form used is $J_0T_2^\ast(L)=\tau_0L^{-\gamma}$, and we report $\gamma$ here.}
	\label{tab:GammaValues}
\end{table*}

\subsection{Two-dimensional arrays} \label{subsec:2DArrays}
\begin{figure*}[t]
	\centering
		\includegraphics{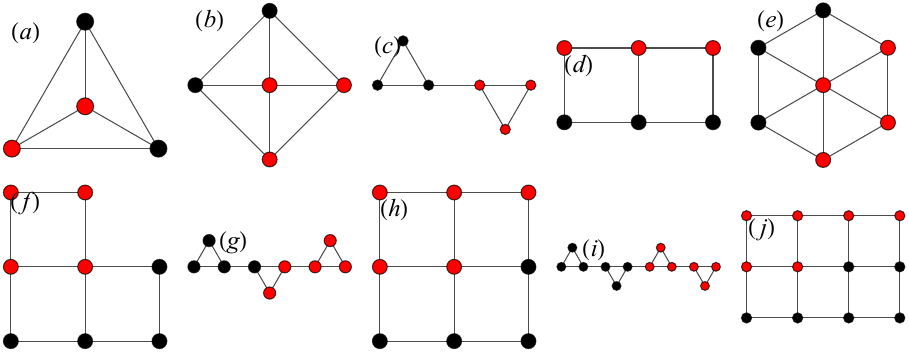}
	\caption{Diagrams of the various two-dimensional arrays considered.  The red dots indicate which spins are assigned to the ``environment'' $B$ in calculating the entanglement entropy.}
	\label{fig:2DArray_Diagrams}
\end{figure*}

We also consider the same problem in two-dimensional (2D) arrays of qubits to further investigate the effects of system geometry on the decoherence time.  We show the multiqubit geometries considered in Fig.~\ref{fig:2DArray_Diagrams}.
We present our results for these geometries in Table \ref{tab:2DArray_T2S}, along with the corresponding results for the line and ring geometries considered earlier.
\begin{table*}
\centering
		\begin{tabular}{|c|c|c|c|c|}
			\hline
			{\bf Geometry} & {$(J_0 T_2^\ast,\alpha)$ ($\sigma_J=0.01J_0$)} & {$(J_0 T_2^\ast,\alpha)$ ($\sigma_J=0.02J_0$)} & {$(J_0 T_2^\ast,\alpha)$ ($\sigma_J=0.05J_0$)} & {$(J_0 T_2^\ast,\alpha)$ ($\sigma_J=0.1J_0$)} \\
			\hline
			\hline
			$L=4$ chain & $(25,0.8)$ & $(16,1.2)$ & $(6.4,1.6)$ & $(3.8,0.6)$ \\
			\hline
			$L=4$ ring & $(35,2)$ & $(21.4,2)$ & $(10,2)$ & $(4.9,1.6)$ \\
			\hline
			$L=4$ decorated triangle [Fig.~\ref{fig:2DArray_Diagrams}(a)] & $(12.5,2)$ & $(5.9,2)$ & $(2.3,2)$ & $(1.2,2)$ \\
			\hline
			\hline
			$L=5$ chain & $(30,2)$ & $(26,0.6)$ & $(6.8,1.35)$ & $(3.2,0.7)$ \\
			\hline
			$L=5$ ring & $(29,0.75)$ & $(11.4,1.8)$ & $(2.5,0.5)$ & $(1.5,0.9)$ \\
			\hline
			$L=5$ decorated square [Fig.~\ref{fig:2DArray_Diagrams}(b)] & $(20,1)$ & $(7.75,1.4)$ & $(2.8,1)$ & $(0.8,0.7)$ \\
			\hline
			\hline
			$L=6$ chain & $(23,0.45)$ & $(8.75,0.6)$ & $(5,1.1)$ & $(1.5,0.7)$ \\
			\hline
			$L=6$ ring & $(30,0.9)$ & $(10.5,1.5)$ & $(10,1)$ & $(3,1.1)$ \\
			\hline
			$L=6$ connected triangles [Fig.~\ref{fig:2DArray_Diagrams}(c)] & $(6.9,1.1)$ & $(4.6,1.6)$ & $(2.55,1.6)$ & $(1.45,1)$ \\
			\hline
			$L=6$ rectangle [Fig.~\ref{fig:2DArray_Diagrams}(d)] & $(17,0.5)$ & $(7.5,0.9)$ & $(2.65,0.9)$ & $(1.35,1)$ \\
			\hline
			\hline
			$L=7$ chain & $(15,0.5)$ & $(6.3,0.3)$ & $(3,0.7)$ & $(1.3,0.7)$ \\
			\hline
			$L=7$ ring & $(11.7,1.3)$ & $(5.8,0.55)$ & $(3.3,1)$ & $(2.35,1.6)$ \\
			\hline
			$L=7$ decorated hexagon [Fig.~\ref{fig:2DArray_Diagrams}(e)] & $(12,0.8)$ & $(7,1)$ & $(1.6,0.6)$ & $(0.6,0.4)$ \\
			\hline
			\hline
			$L=8$ chain & $(9,0.5)$ & $(4,0.5)$ & $(2.7,0.7)$ & $(1.2,0.7)$ \\
			\hline
			$L=8$ ring & $(12,0.5)$ & $(11,0.55)$ & $(5.3,0.6)$ & $(2.9,1)$ \\
			\hline
			$L=8$ near square [Fig.~\ref{fig:2DArray_Diagrams}(f)] & $(8.4,1)$ & $(4.25,0.8)$ & $(1.4,0.8)$ & $(0.95,0.8)$ \\
			\hline
			\hline
			$L=9$ chain & $(8,0.7)$ & $(7.5,0.9)$ & $(4.75,1)$ & $(1.8,0.75)$ \\
			\hline
			$L=9$ ring & $(1.8,0.4)$ & $(1.45,0.4)$ & $(0.9,0.45)$ & $(0.7,0.7)$ \\
			\hline
			$L=9$ connected triangles [Fig.~\ref{fig:2DArray_Diagrams}(g)] & $(4.8,1)$ & $(3.3,1)$ & $(1.6,1)$ & $(0.45,0.7)$ \\
			\hline
			$L=9$ square [Fig.~\ref{fig:2DArray_Diagrams}(h)] & $(13.1,0.78)$ & $(7.9,1.05)$ & $(3.5,1)$ & $(2.3,0.7)$ \\
			\hline
			\hline
			$L=12$ chain & $(2.4,0.4)$ & $(0.4,0.3)$ & $(0.3,0.35)$ & $(0.1,0.65)$ \\
			\hline
			$L=12$ ring & $(10,1.05)$ & $(8,0.9)$ & $(5.25,0.9)$ & $(2.6,0.7)$ \\
			\hline
			$L=12$ connected triangles [Fig.~\ref{fig:2DArray_Diagrams}(i)] & $(3.5,1)$ & $(1.25,1)$ & $(0.29,1)$ & $(0.25,0.7)$ \\
			\hline
			$L=12$ rectangle [Fig.~\ref{fig:2DArray_Diagrams}(j)] & $(4.75,0.8)$ & $(1.25,1)$ & $(0.65,0.65)$ & $(0.35,0.6)$ \\
			\hline
		\end{tabular}
	\caption{$T_2^\ast$ and $\alpha$ for the geometries shown in Fig.~\ref{fig:2DArray_Diagrams}, along with the corresponding results for the chain and ring geometries, for $\sigma_J=0.1J_0$.}
	\label{tab:2DArray_T2S}
\end{table*}
We see from Table \ref{tab:2DArray_T2S} that geometry can have a significant effect on the decoherence time.  As an extreme example, we see that the geometry of the $L=12$ case can change $T_2^\ast$ by an order of magnitude, with $J_0T_2^\ast=0.1$ for a chain but $J_0T_2^\ast=2.6$ for a ring, both for $\sigma_J=0.1J_0$.

We also record the values of $\alpha$ found for each case.  We see that, for small systems (at most four spins), $\alpha=2$.  Therefore, the decay of the return probability to its steady-state value is Gaussian, in agreement with previous work.  However, for larger systems, we find smaller values of $\alpha$, thus indicating that the decay is no longer Gaussian for these system sizes.  We see from the plots in Fig.~\ref{fig:AlphaPlots} that there is a downward trend in $\alpha$ as the system size increases for both the chain and ring geometries.

\section{Analysis} \label{sec:Analysis}
We now turn to providing an explanation for the behavior of $T_2^\ast$ and $\alpha$ that we find.  The general trend that we note is that the decay tends to be Gaussian for small system sizes and lower values of $\sigma_J$, while it deviates from Gaussian decay otherwise.  We now show that we can explain this behavior using perturbation theory.  We begin by demonstrating that the return probability, to first order in perturbation theory in the noise-induced deviations of the exchange couplings from their intended values, can be written as the sum of sinusoidal oscillations with Gaussian decay.  We define these deviations as $\Delta_{m,m+1}=J_{m,m+1}-J_0$.

We start by writing down the general form of the wave function to first order:
\begin{equation}
	\ket{\psi_n^1}=\sum_{k\neq n}\frac{M^k_n(\{\Delta_{m,m+1}\})}{E_n^0-E_k^0}\ket{\psi_k^0},
\end{equation}
where $M^k_n(\{\Delta_{m,m+1}\})$ is a linear function of $\Delta_{m,m+1}$,
\begin{equation}
	M^k_n(\{\Delta_{m,m+1}\})=\sum_m M^k_{n,m}\Delta_{m,m+1},
\end{equation}
$\ket{\psi_k^0}$ are the unperturbed eigenstates, and $E_k^0$ are the corresponding energies.  The perturbed energies are given by
\begin{equation}
	E_n^1=E_n^0+F_n(\{\Delta_{m,m+1}\}),
\end{equation}
where $F_n(\{\Delta_{m,m+1}\})$ is again a linear function of $\Delta_{m,m+1}$,
\begin{equation}
	F_n(\{\Delta_{m,m+1}\})=\sum_m F_{n,m}\Delta_{m,m+1}.
\end{equation}
We now let $\ket{\Psi_0}$ be the initial state of the system.  We can then write the state of the system $\ket{\Psi(t)}$ at any time as
\begin{equation}
	\ket{\Psi(t)}=\sum_n a_ne^{-iE_n^1t/\hbar}\ket{\psi_n},
\end{equation}
where $a_n=\braket{\psi_n|\Psi_0}$.  From this, we can determine the return probability $P_R(t)$ to first order in the $\Delta_{m,m+1}$:
\begin{widetext}
	\begin{eqnarray*}
		P_R(t)&=&\left |\braket{\Psi_0|\Psi(t)}\right |^2 \cr
		&=&\sum_{n,p}\left\{\left |a_n^0\right |^2\left |a_p^0\right |^2+2\left |a_n^0\right |^2\sum_{k\neq p}\left [\frac{(\mbox{Re}\, M^k_{p,m})\Delta_{m,m+1}}{E_p^0-E_k^0}a_k^0(a_p^0)^\ast+(n\leftrightarrow p)\right ]\right\}e^{-i(E_n^1-E_p^1)t/\hbar},
	\end{eqnarray*}
	where $a_k^0=\braket{\psi_k^0|\Psi_0}$.  We now average all $\Delta_{m,m+1}$ over the Gaussian distribution,
	\begin{equation}
		f_\Delta(\Delta)=\frac{1}{\sigma_J\sqrt{2\pi}}e^{-\Delta^2/2\sigma_J^2}.
	\end{equation}
	After doing so, we obtain
	\begin{equation}
		\braket{P_R(t)}=\sum_{n,p}\left |a_n^0\right |^2\left |a_p^0\right |^2\exp\left [-\tfrac{1}{2}\sigma_J^2\sum_j\left (\frac{F_{n,j}-F_{p,j}}{\hbar}\right )^2t^2\right ]\cos\left (\frac{E_n^0-E_p^0}{\hbar}t\right ).
	\end{equation}
\end{widetext}
We thus see that, as stated earlier, the expression obtained to first order in perturbation theory, after averaging over realizations of noise, is simply a sum of sinusoidal oscillations with Gaussian decays.  We now make use of this result to explain how one may see deviations from an overall Gaussian decay of the return probability.  Assuming that perturbation theory holds, which it will if $\sigma_J$ is much smaller than the smallest (unperturbed) energy level spacing, we may obtain a non-Gaussian decay if there is a large number of these terms in the expression.  For smaller system sizes, we see only a small number of terms, and thus we may obtain an approximately Gaussian overall decay profile, especially if one of the Gaussians decays over a much longer time scale than the others.  For larger systems, however, there will be a large number of these terms, and thus we will obtain significant deviations from a Gaussian profile, even if perturbation theory holds.  This is consistent with the trends that we obtained.

The second way in which the decay may be non-Gaussian is if perturbation theory simply breaks down.  As mentioned earlier, perturbation theory holds only if $\sigma_J$ is much smaller than the smallest energy level spacing.  This fails to hold if either $\sigma_J$ becomes too large or if the system size $L$ is too large.  We now show through examples that perturbation theory fails for the system sizes and values of $\sigma_J$ for which we find non-Gaussian decay.

As our first example, let us consider the four-spin ring.  In this case, we find that the smallest level spacing in the system is $4J_0$, which is much larger than any of the $\sigma_J$ values considered in this work, so we expect perturbation theory to work very well in this case.  We also find that, in the expression for $P_R(t)$, one term will dominate the long-time behavior of the system:
\begin{eqnarray}
	P_R(t)&=&\tfrac{7}{18}+\tfrac{1}{3}e^{-2\sigma_J^2 t^2}\cos{4J_0t}+\tfrac{1}{6}e^{-8\sigma_J^2t^2}\cos{8J_0t} \cr
	&+&\tfrac{1}{9}e^{-18\sigma_J^2t^2}\cos{12J_0t}.
\end{eqnarray}
This expression, provided here for the sake of a self-contained presentation, is consistent with that found in Ref.\ \onlinecite{PhysRevB.103.205402} (note again the different definition of the exchange coupling compared to Ref.\ \onlinecite{PhysRevB.103.205402}).  We find that the decay is Gaussian in this case for all values of $\sigma_J$ except for $\sigma_J=0.1J_0$.  This deviation is likely due to the fact that the time scale of the decay is becoming comparable to the period of the oscillations.

Let us now consider the four-spin chain, where we see that the decay fails to be Gaussian, even for $\sigma_J=0.01J_0$.  We find that the smallest level spacing is $1.17157J_0$, so we still expect perturbation theory to hold for all values of $\sigma_J$ that we consider.  However, the expression for $P_R(t)$ contains a much larger number of terms than for the ring geometry.  As a result, the overall decay profile will be non-Gaussian.

\begin{figure*}[th]
	\centering
	\includegraphics[width=0.32\linewidth]{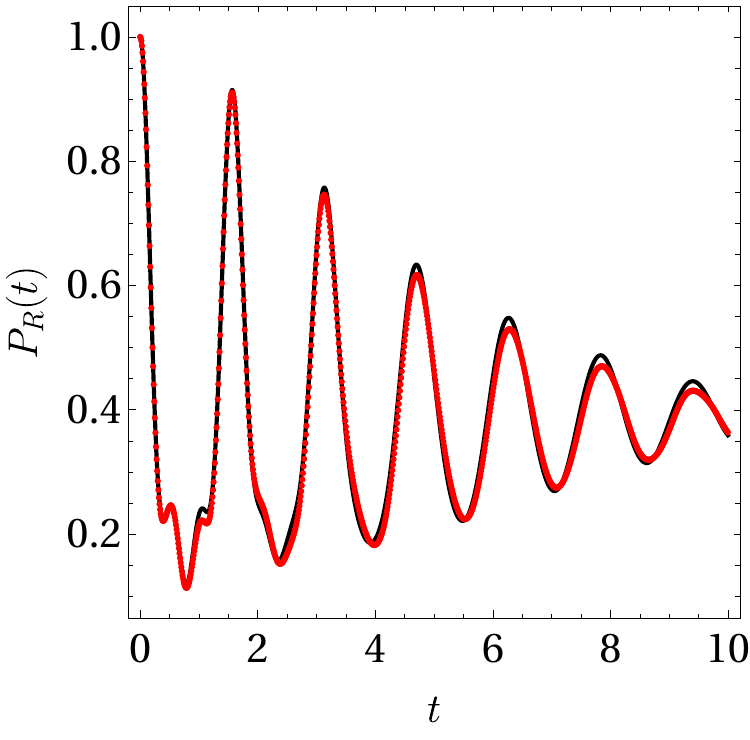}
	\includegraphics[width=0.32\linewidth]{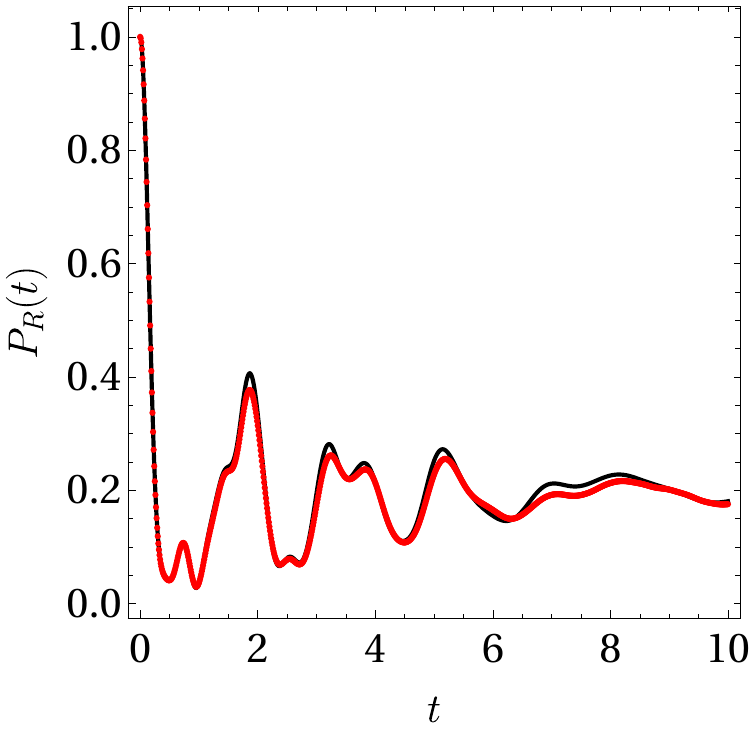}
	\includegraphics[width=0.32\linewidth]{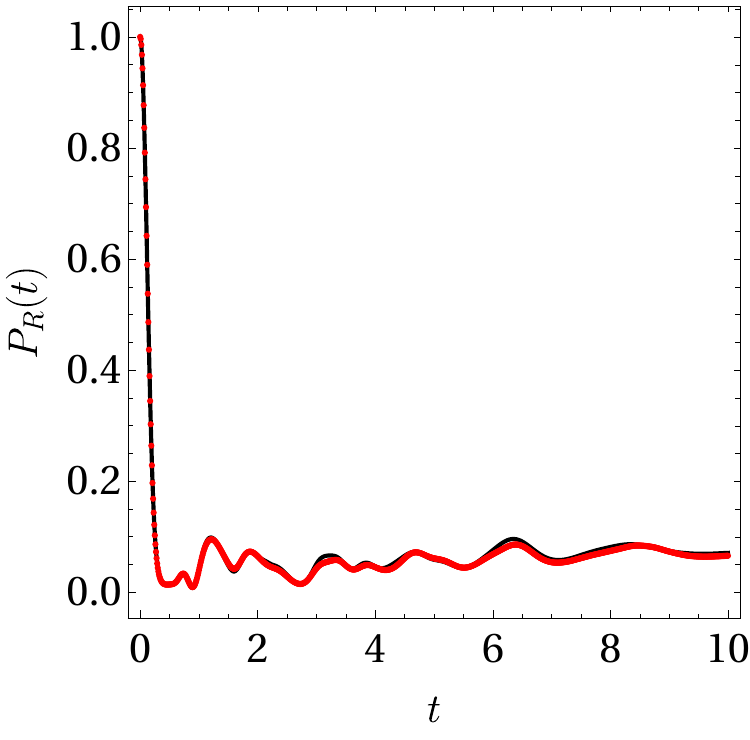}
	\caption{Plot of both exact return probability (red dots) and the perturbative approximation (black curve) as a function of time and for $\sigma_J=0.1J_0$ for a $4$-spin ring (left), a $7$-spin chain (center), and a $10$-spin chain (right).}
	\label{fig:RP_PerturbVsExact}
\end{figure*}

We now consider the second factor, namely, that perturbation theory breaks down for large system sizes or for large enough $\sigma_J$.  We consider three cases: the $4$-spin ring, and the $7$- and $10$-spin chains.  We plot both the exact and perturbative results for the return probability, all for $\sigma_J=0.1J_0$, in Fig.~\ref{fig:RP_PerturbVsExact}.
We see that perturbation theory works surprisingly well, even for the larger systems; while there is a noticeable deviation from the exact result for $L=7$ and $10$, it is not very large.  We thus conclude that the proliferation of terms in the perturbative result is by far the dominant factor in deciding whether the decay is Gaussian or not.  Physically, this is related to the fact that, as the system size increases, the number of energy eigenstates in the $S_z=0$ (even $L$) or $S_z=-\tfrac{1}{2}$ (odd $L$) subspace grows as $2^L/\sqrt{L}$, and thus, the number of possible transitions from one energy level to another also grows rapidly.

\section{Average spin, Hamming distance, and entanglement entropy} \label{sec:OtherQs}
We now consider three other measures of the system: the average spin, the Hamming distance, and the entanglement entropy.  The first two can be measured experimentally, but the entanglement entropy is of theoretical interest, and thu,s we include it as well.  All results presented as examples in this section are for $\sigma_J=0.1J_0$ and for $L=8$.  We first determine the expectation value of the $z$ component of one of the spins, $\braket{\sigma_{1,z}}(t)$; in the chain geometry, we use the leftmost spin, while we choose a spin that is pointing down in the ring geometry.

We then determine the Hamming distance, which is a measure of how different the system's state is from its initial state; in general, it is the number of elements in a string that differ from those of a given reference string.  In our case, it is related to the number of qubits that are measured to be in a different state than they started in.  We use a normalized Hamming distance, which we will denote $D(t)$, given for the initial state that we use by
\begin{equation}
D(t)=\tfrac{1}{2}-\frac{1}{2L}\sum_{k=1}^{L}(-1)^k\braket{\sigma_k}.
\end{equation}
This definition is normalized in such a way that $D(t)=0$ if the current state is exactly the initial state $\ket{\Psi_0}$ and $D(t)=1$ if all qubits are in the state opposite their initial state, i.e., if $\ket{\Psi(t)}=\ket{\uparrow\downarrow\uparrow\cdots}$.

Finally, we determine the entanglement entropy, which is of theoretical interest, but is very difficult to measure experimentally.  We calculate the entanglement entropy as follows.  We first form the density matrix for the current state of the system, $\rho(t)=\ket{\Psi(t)}\bra{\Psi(t)}$.  We then divide the system into a subsystem, $A$, and the ``environment,'' $B$.  In the chain geometry, we choose $A$ to be the left half of the system and $B$ to be the right half; in the case of odd $L$, the middle spin is assigned to $B$.  For the nontrivial $2$D arrays in Fig.\ \ref{fig:2DArray_Diagrams}, we indicate the environment with red dots.  In all cases, the smaller of the two systems in the case of odd $L$ will be subsystem $A$.  By doing this, we form the reduced density matrix by tracing out $B$, i.e., $\rho_A(t)=\mbox{Tr}_B\,\rho(t)$.  The entanglement entropy is then given by
\begin{equation}
S(t)=-\mbox{Tr}_A[\rho_A(t)\ln\rho_A(t)].
\end{equation}

\begin{figure*}
	\centering
		\includegraphics[width=0.32\textwidth]{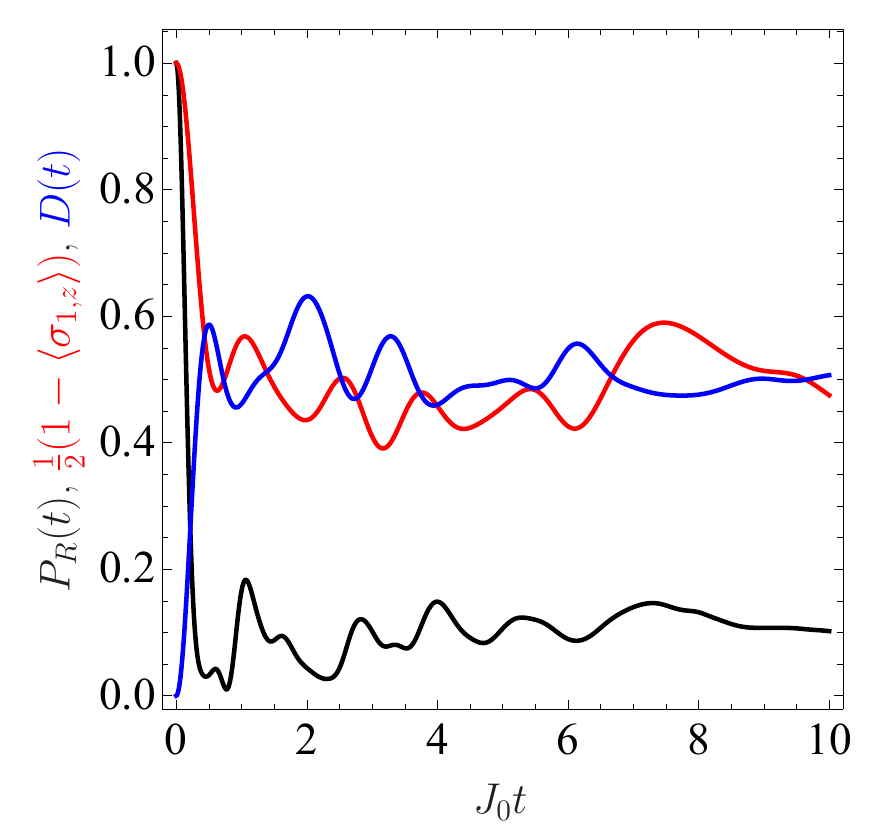}
		\includegraphics[width=0.32\textwidth]{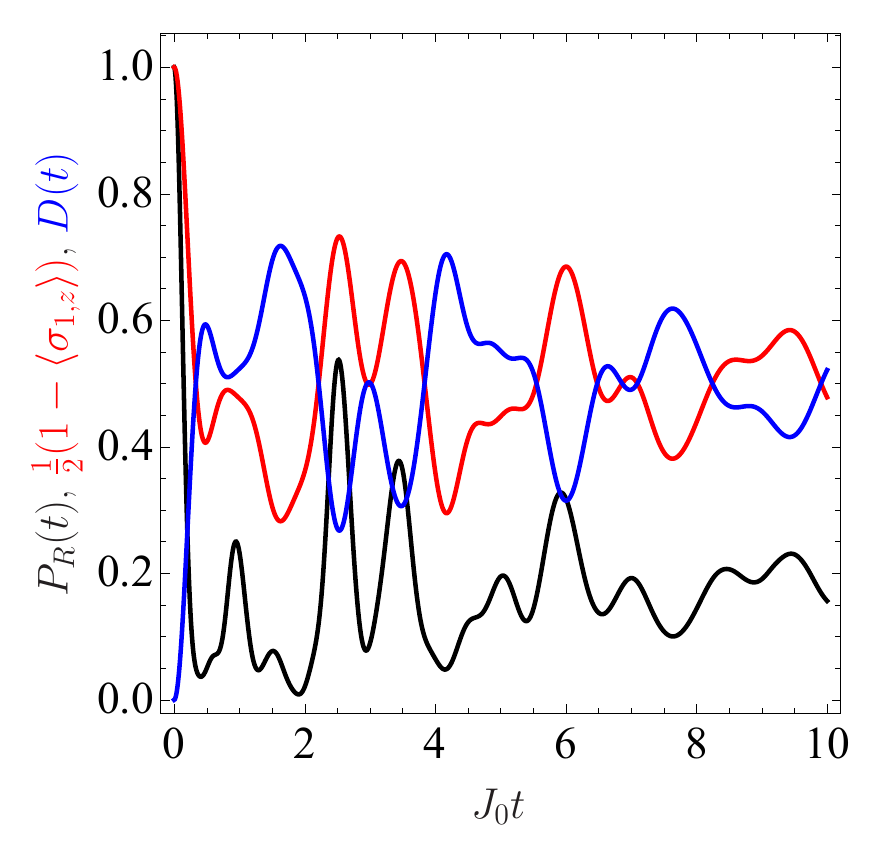}
		\includegraphics[width=0.32\textwidth]{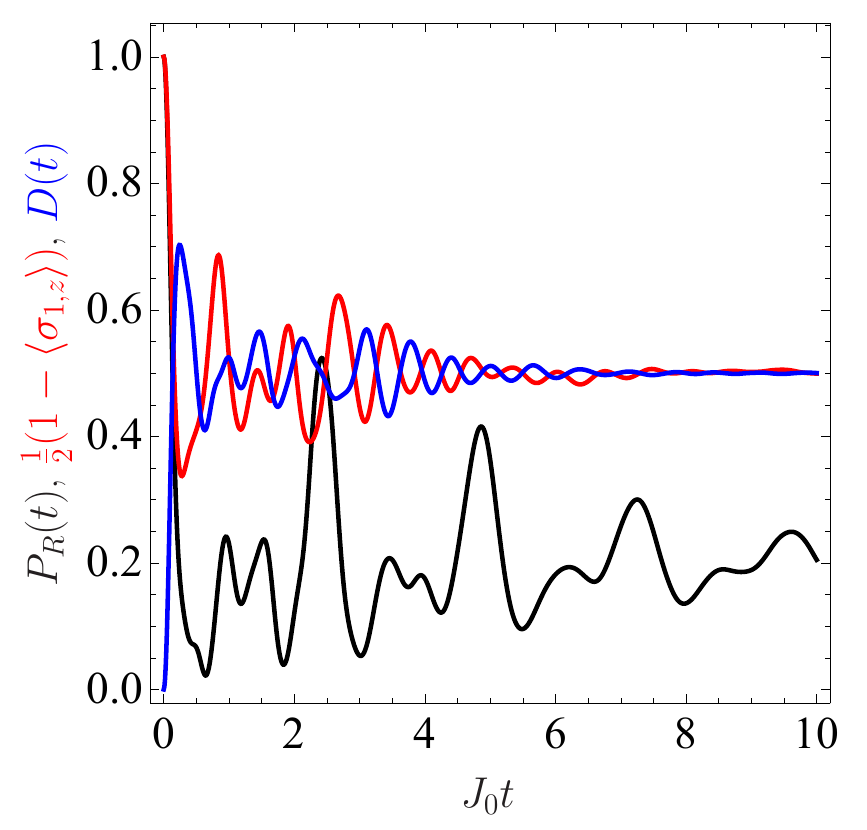}
	\caption{Plot of the return probability (black), average spin (red), and Hamming distance (blue) for an eight-spin chain (left), an eight-spin ring (center), and eight spins arranged as shown in Fig.~\ref{fig:2DArray_Diagrams}(f) (right) for $\sigma_J=0.1J_0$.}
	\label{fig:CombPlot_Example}
\end{figure*}

\begin{figure*}
	\centering
		\includegraphics[width=0.32\textwidth]{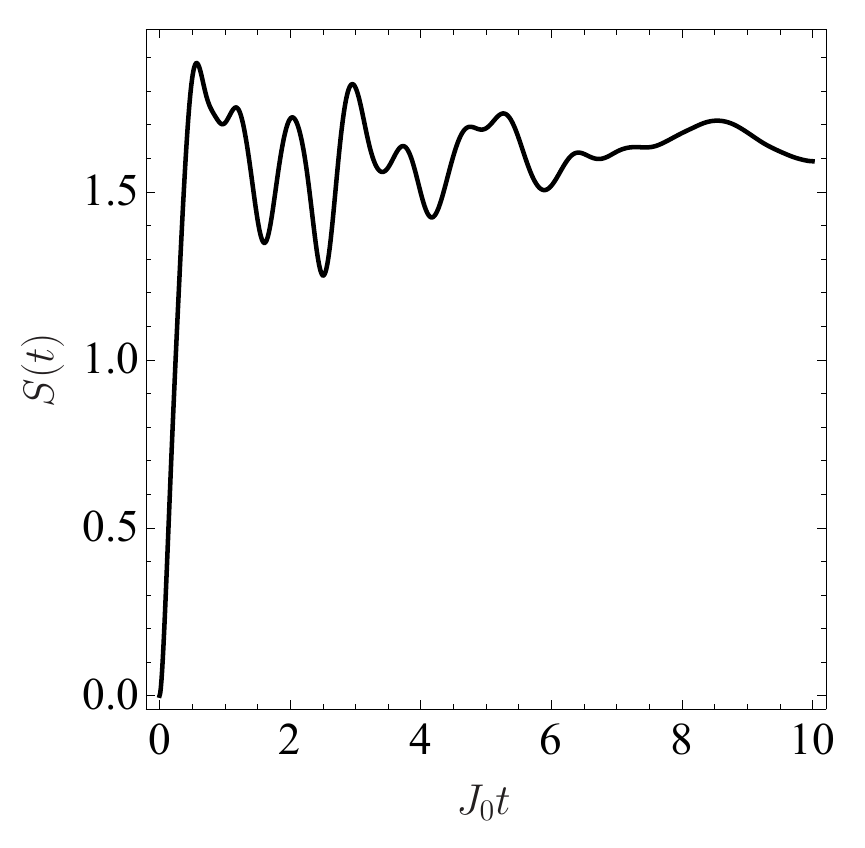}
		\includegraphics[width=0.32\textwidth]{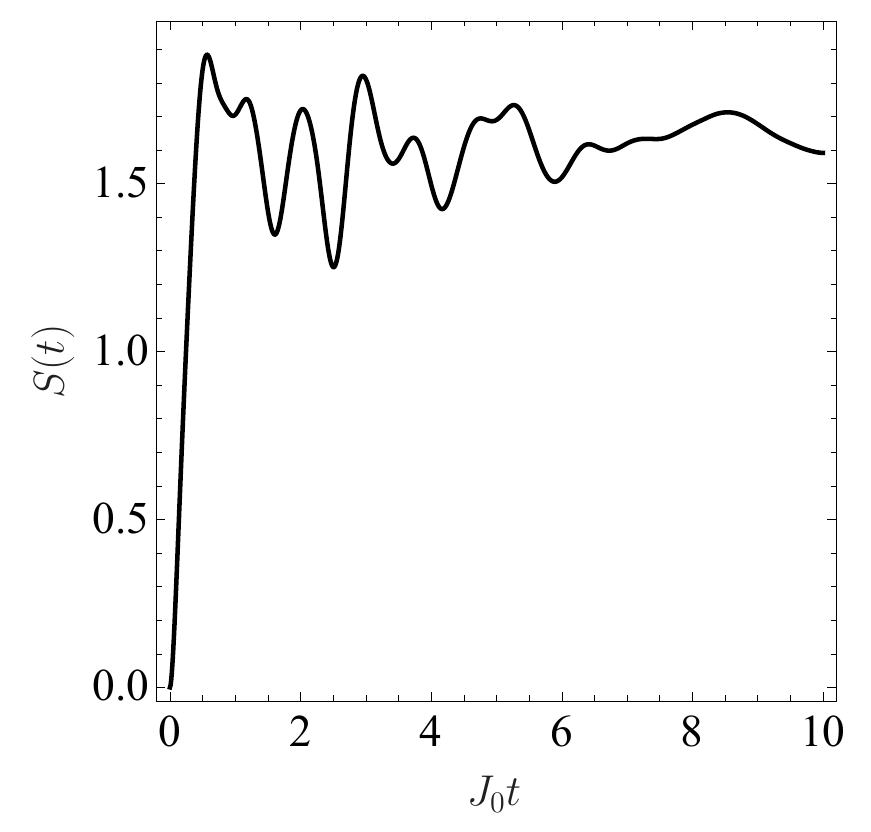}
		\includegraphics[width=0.32\textwidth]{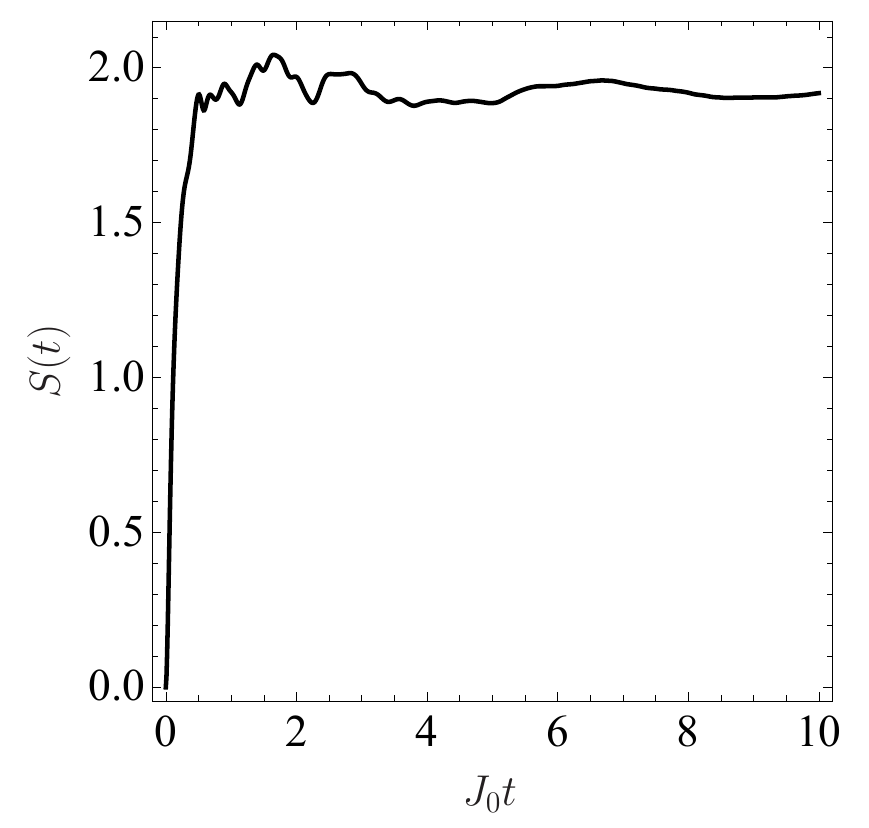}
	\caption{Plot of the entanglement entropy for an eight-spin chain (left), an eight-spin ring (center), and eight spins arranged as shown in Fig.\ \ref{fig:2DArray_Diagrams}(f) (right) for $\sigma_J=0.1J_0$.}
	\label{fig:EEPlot_Example}
\end{figure*}

We show plots of the return probability, average spin, and Hamming distance for $L=8$ for the chain and ring geometries, as well as the array shown in Fig.~\ref{fig:2DArray_Diagrams}(f), in Fig.~\ref{fig:CombPlot_Example}.
We see that the average spin and Hamming distance are good proxies for the return probability for measuring the decoherence time $T_2^\ast$.  We note that, in the ring geometry, the average spin and Hamming distance are related by $D(t)=\tfrac{1}{2}\left (\braket{\sigma_{1,z}}+1\right )$; this is due to the symmetry of both the underlying system and the initial condition.  We present the corresponding results for the entanglement entropy in Fig.~\ref{fig:EEPlot_Example}.

\begin{figure}
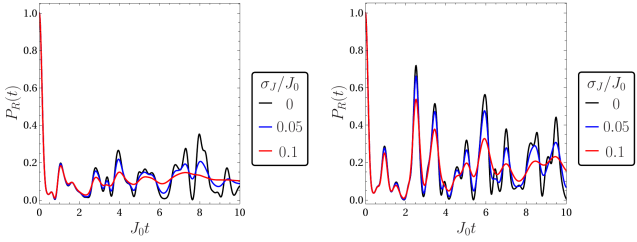

	\centering
		\includegraphics[width=0.49\columnwidth]{RP_DifferentSigmaJ_8Spins.pdf}
		\includegraphics[width=0.49\columnwidth]{RP_DifferentSigmaJ_8Spins_Ring.pdf}
	\caption{Plot of the return probability $P_R(t)$ for an eight-spin chain (left) and an eight-spin ring (right) for three different values of $\sigma_J$.}
	\label{fig:RP_DifferentSigmaJ}
\end{figure}

\begin{figure}
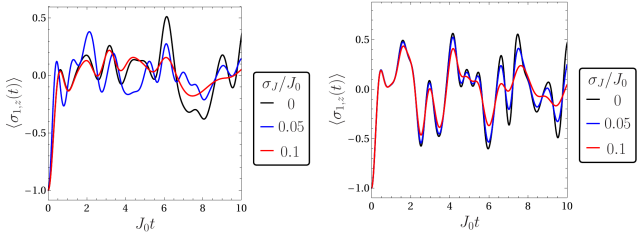

	\centering
		\includegraphics[width=0.49\columnwidth]{SP_DifferentSigmaJ_8Spins.pdf}
		\includegraphics[width=0.49\columnwidth]{SP_DifferentSigmaJ_8Spins_Ring.pdf}
	\caption{Plot of the expectation value of the leftmost spin $\braket{\sigma_{1,z}(t)}$ for an eight-spin chain (left) and an eight-spin ring (right) for three different values of $\sigma_J$.}
	\label{fig:SP_DifferentSigmaJ}
\end{figure}

\begin{figure}
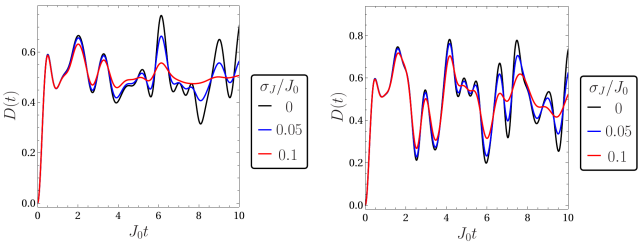

	\centering
		\includegraphics[width=0.49\columnwidth]{HD_DifferentSigmaJ_8Spins.pdf}
		\includegraphics[width=0.49\columnwidth]{HD_DifferentSigmaJ_8Spins_Ring.pdf}
	\caption{Plot of the expectation value of the Hamming distance $D(t)$ for an eight-spin chain (left) and an eight-spin ring (right) for three different values of $\sigma_J$.}
	\label{fig:HD_DifferentSigmaJ}
\end{figure}

\begin{figure}
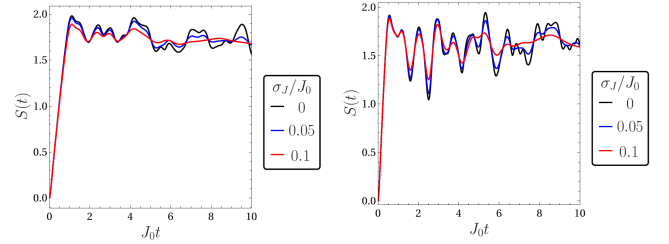

	\centering
		\includegraphics[width=0.49\columnwidth]{EE_DifferentSigmaJ_8Spins.pdf}
		\includegraphics[width=0.49\columnwidth]{EE_DifferentSigmaJ_8Spins_Ring.pdf}
	\caption{Plot of the expectation value of the entanglement entropy $S(t)$ for an eight-spin chain (left) and an eight-spin ring (right) for three different values of $\sigma_J$.}
	\label{fig:EE_DifferentSigmaJ}
\end{figure}

We additionally provide plots for $L=8$ for both the chain and ring geometries of all four of these quantities for three different values of $\sigma_J$ in Figs.\ \ref{fig:RP_DifferentSigmaJ}--\ref{fig:EE_DifferentSigmaJ}.

\section{Conclusion} \label{sec:Conclusion}
We studied the decoherence time $T_2^\ast$ for arrays of spins of varying sizes $L$ and of different geometries in the presence of quasistatic charge noise.  We determined $T_2^\ast$ by numerically calculating the return probability of the system as a function of time and visually finding the curve of the form [Eq.~\eqref{Eq:EnvFittingForm}] that best fits the envelope of the return probability.  In all cases, we used an alternating arrangement of spins, $\ket{\Psi_0}=\ket{\downarrow\uparrow\downarrow\cdots}$, as our initial state.  We did this for chains and rings of different lengths $L=3$--$14$, as well as for a number of two-dimensional arrays, illustrated in Fig.~\ref{fig:2DArray_Diagrams}.  We showed that $T_2^\ast$ follows a power law as a function of $L$ for both the chain and ring geometries, i.e., $T_2^\ast\propto L^{-\gamma}$.  We found that the exponent $\gamma$ varies with the amount of charge noise present in the system, as measured by $\sigma_J$, and also depends on the geometry (chain or ring).  We found that $\gamma$ decreases with increasing $\sigma_J$ for the ring geometry but tends to increase with increasing $\sigma_J$ for the chain geometry.  We also illustrated through a number of examples that the geometry of the system can greatly affect $T_2^\ast$, sometimes by an order of magnitude.  We also recorded the values of $\alpha$ extracted from this fit; we found that, for small systems (at most four spins), $\alpha=2$, and thus, the decay is Gaussian, in agreement with previous work.  For larger system sizes, however, we found smaller values of $\alpha$, meaning that the decay ceases to be Gaussian for larger systems.

We showed that our results for $T_2^\ast$ and $\alpha$ could be connected with perturbation theory and its breakdown.  We showed that, within first-order perturbation theory in the deviation of the exchange couplings from their intended values, the return probability could be written as a sum of sinusoidal oscillations with Gaussian decay profiles.  We then argued that an overall non-Gaussian decay profile can emerge either due to the return probability having a sufficiently large number of these terms, or as a result of perturbation theory simply breaking down.  Both of these tend to happen for large system sizes or, in the latter case, for sufficiently strong noise, consistent with our numerical results.  We saw, however, that the proliferation of terms in the expression for the return probability is by far the dominant factor, as perturbation theory turns out to work surprisingly well, even for larger systems in which we expect it to fail.  This large number of terms is ultimately due to the fact that, as the system size increases, the number of energy eigenstates within the $S_z=0$ (even $L$) or $S_z=-\tfrac{1}{2}$ (odd $L$) subspace grows exponentially.  This means that there is a larger number of transitions that a perturbation can induce between two eigenstates, which means that the perturbative expansion will have a larger number of terms.

We also determined the average spin, the Hamming distance, and the entanglement entropy.  We saw that these three quantities tend to a steady-state value over a time scale similar to that of the return probability.  The entanglement entropy is solely of theoretical interest, as it cannot be measured experimentally, but the other two quantities, the spin expectation value and Hamming distance, can.  As a result, the spin expectation value and the Hamming distance are useful proxies for determining $T_2^\ast$.

The investigation of how $T_2^\ast$ depends on the system size and geometry (i.e., arrangement of spins) is critical because decoherence has a detrimental effect on the fidelity of gates performed on a system of qubits and because, eventually, systems of many qubits will need to be built in order to create a working quantum computer.  In addition to noise as a source of decoherence, which would be present even for a single qubit, interactions among qubits, while necessary for performing two-qubit gates, can also be a source of decoherence via crosstalk among the qubits.  Furthermore, how the qubits are arranged can have an effect on the amount of crosstalk that a given qubit experiences simply due to the different connectivities of the qubits with one another in different geometries.  Therefore, studying the effects of system geometry is just as important as studying the effect of system size.

We should note that we assumed that the system is simply left to evolve on its own from the initial state $\ket{\Psi_0}$; we applied no pulses to the system to perform gates.  Therefore, the $T_2^\ast$ values that we found are without special pulse sequences intended to combat decoherence, such as the Hahn echo and CPMG sequences.  Use of such sequences in experiments can therefore effectively extend the decoherence time.  We also note that, since we did not consider the effects of Overhauser noise in our work, our results are applicable specifically to Si-based systems.  Overhauser noise will reduce the $T_2^\ast$ values found here further in GaAs, where it is unavoidable due to the fact that, unlike Si, no stable non-magnetic isotopes of Ga or As exist, and thus isotopic purification aimed at removing such isotopes is impossible in GaAs.

\acknowledgments
This work was supported by the Laboratory for Physical Sciences.  The authors acknowledge the University of Maryland supercomputing resources made available for conducting the research reported in this paper.

\bibliography{DecoherenceWithL_SpinQubits}

\end{document}